\documentclass[12pt, twoside, leqno]{article}

\usepackage{amsmath,amsthm}
\usepackage{amssymb}

\usepackage{enumerate}

\newtheorem{thm}{Theorem}[section]

\newtheorem{lem}[thm]{Lemma}

\theoremstyle{definition}

\newtheorem{rem}[thm]{Remark}

\numberwithin{equation}{section}

\begin{document}


\baselineskip=17pt


\title{Optimizing expected utility of dividend payments for a Cram\'er-Lundberg risk process}

\author{Sebastian Baran\\
Department of Mathematics\\
Cracow University of Economics\\
31-510 Cracow, Poland\\
E-mail: sebastianbaran13@gmail.com
\and
Zbigniew Palmowski\\
Faculty of Pure and Applied Mathematics\\
Wroc{\l}aw University of Science and Technology\\
50-370 Wroc{\l}aw, Poland\\
E-mail: zbigniew.palmowski@gmail.com}

\date{}

\maketitle


\renewcommand{\thefootnote}{}



\renewcommand{\thefootnote}{\arabic{footnote}}
\setcounter{footnote}{0}


\begin{abstract}
We consider the problem of maximizing  the discounted utility of dividend payments of an insurance company whose reserves are modeled as a classical Cram\'er-Lundberg risk process. We investigate this optimization problem under the constraint that dividend rate is bounded. We prove that the value function fulfills the Hamilton-Jacobi-Bellman equation and we identify the optimal dividend strategy.
\end{abstract}

\section{Introduction}\label{sec:Introduction}
The problem of finding optimal dividend strategies for an insurance company has been studied since De Finetti \cite{DeFin}.
The continuous time risk process was studied for the first time in a seminal paper of Gerber \cite{Gerb}.
He assumed that the reserve process $R=(R_t)_{t\geq 0}$ of an insurance company is modeled by a classical Cram\'er-Lundberg risk process:
\begin{equation}
R_t=x+\mu t-\sum_{i=1}^{N_t}Y_i,\label{CL}
\end{equation}
where $Y_1,Y_2,\ldots$ are i.i.d positive random variables with absolutely continuous d.f. $F_Y$ representing the claims, $N=(N_t)_{t\geq 0}$ is an independent Poisson process with intensity $\lambda>0$ modeling the times at which the claims occur, $x\geq 0$ denotes the initial surplus and $\mu$ is a premium intensity.
We define above Poisson process $N$ and the sequence $\{Y_i, i\in \mathbf{N}\}$ on a common probability space $(\Omega, \mathcal{F}, \mathbb{P})$.

For the classical dividend problem, apart of the reserve process (\ref{CL}), we consider the dividend payments.
Let $C=(C_t)_{t\geq 0}$ be an adapted and nondecreasing process representing all accumulated dividend payments up to time $t$.
Then the regulated process $X=(X_t)_{t\geq 0}$ is given by:
\begin{equation}\label{regproc}
X_t=R_t-C_t.
\end{equation}
We observe the regulated process $X_t$ until the time of ruin:
\begin{equation*}
\tau=\inf\{t\geq 0\colon X_t<0\}.
\end{equation*}
Obviously the time of the ruin of an insurance company depend on dividend strategy and after ruin occurs no dividends are paid. As usual we assume that net profit condition $\mu>\lambda E(Y_1)$ for Cram\'er-Lundberg risk process is fulfilled.

For the classical dividend problem we define the value of the dividends as:
\begin{equation*}
v_C(x)=\mathbb{E}_x\left(\int_0^{\tau}e^{-\beta t}dC_t\right),
\end{equation*}
where $\beta > 0$ is a discount factor, $\mathbb{E}_x$ means expectation with respect to $\mathbb{P}_x(\cdot)=P(\cdot|X_0=x)$ and the value function as:
\begin{equation*}
v(x)=\sup_{C\in\mathrm{C}}v_C(x),
\end{equation*}
for $\mathrm{C}$ being the set of all admissible accumulated dividend strategies $(C_t)_{t\geq 0}$.

In the mathematical finance and actuarial literature, there is a good deal of work being done
on dividend barrier models and the problem of finding an optimal policy of paying
dividends, see Schmidli \cite{Schmidli} for an overview. In this paper we
assume additionally  that $(C_t)_{t\geq 0}$ is absolutely continuous with respect to the Lebesgue measure; see e.g.
Hubalek and Schachermayer \cite{Hub}.
Then the process $C$ admits a density process denoted by $c=(c_t)_{t\geq 0}$
modeling the intensity of the dividend payments in continuous time. That is, for $t\geq 0$:
\begin{equation*}
C_t=\int_0^t c_sds\ \textrm{a.s.}
\end{equation*}
Then we can consider the discounted cumulative utility of dividend payments:
\begin{equation}\label{Def:DividendsValue}
v_c(x)=\mathbb{E}_x\left(\int_0^{\tau}e^{-\beta t}U(c_t)dt\right),
\end{equation}
where $U$ is some fixed utility function, and the value function equals:
\begin{equation}\label{Def:ValueFunction}
v(x)=\sup_{c\in\mathfrak{C}}v_c(x),
\end{equation}
for $\mathfrak{C}$ being the set of all admissible strategies $(c_t)_{t\geq 0}$.

We assume that dividend density process $(c_t)_{t\geq 0}$ is admissible
if it is a nonnegative, adapted and c\`adl\`ag process and there is no dividend after ruin occurs:  $c_t=0$ for all $t\geq \tau$. As noted in \cite[Rem. 2.1]{Schmidli} we choose c\`adl\`ag strategies instead of more often used left-continuous ones. We study this optimization problem under the constraint that only dividend strategies with dividend rate bounded by a fixed constant are admissible, i.e. $0\leq c_t\leq c_0<\infty$ for all $t\geq 0$ and
some fixed number $c_0>0$. Finally, we assume that the ruin cannot be caused by the dividend payment.

Under this set-up we prove that value function is differentiable and it solves the Hamilton-Jacobi-Bellman (HJB) equation.

\section{Some properties of the Value Function}\label{sec:VF}

Recall that the regulated process is given by:
\begin{equation*}
X_t=x+\mu t-\sum_{i=1}^{N_t}Y_i-\int_0^t c_sds
\end{equation*}
and the value function are defined as in \eqref{Def:DividendsValue}.

From now on we will assume that $U:\mathbb{R}_{\geq 0}\to\mathbb{R}_{\geq 0}$ is continuous, nonnegative, strictly increasing, strictly concave and $U(0)=0$.

\begin{lem}\label{lem:AC}
The optimal value function $v(x)$ is nonnegative and bounded by $\frac{U(c_0)}{\beta}$ and converges to $\frac{U(c_0)}{\beta}$ a.s. as $x\to\infty$.
Furthermore, $v(x)$ is an increasing, Lipschitz continuous and therefore absolutely continuous.
\end{lem}
\begin{proof}
Using the same strategy for two initial capitals shows that $v(x)$ is increasing. Since $U$ is nonnegative hence $v$ is nonnegative. Clearly, $v(x)\leq\int\limits_0^{\infty}U(c_0)e^{-\beta t}dt=\frac{U(c_0)}{\beta}$. Consider the strategy $c^*=(c_t^*)$ which pay dividends at constant rate $c_0$ for all $t\geq 0$ and associated with strategy $c^*$ and initial capital $x$ ruin time $\tau^{c^*}_x$. Then:
\begin{equation}\label{ineq:NeededForLimitOfV}
v(x)\geq\mathbb{E}_x\left(\int_0^{\tau^{c^*}_x}e^{-\beta t}U(c_0)dt\right)=\left(1-\mathbb{E}_x \left[e^{-\beta\tau^{c^*}_x}\right]\right)\frac{U(c_0)}{\beta}.
\end{equation}

We will show now that the ruin time $\tau^{c^*}_x$ converges to infinity almost surely as $x\to\infty$. Choose $y\geq x$ and one realization $\omega\in \Omega$
of $\sum_{i=1}^{N_t}Y_i$. We will denote it by adding superscript $\omega$ to the respective counterparts. Since
\begin{equation*}
y+(\mu-c_0)t-\sum_{i=1}^{N^\omega_t}Y^\omega_i \geq x+(\mu-c_0)t-\sum_{i=1}^{N_t^\omega}Y^\omega_i
\end{equation*}
it follows that $\tau^{c^*}_y\geq \tau^{c^*}_x$ i.e. the function $x\rightarrow \tau^{c^*}_x$ is nondecreasing for fixed $\omega$. 
Then $\tau^{c^*}_x$ has either finite or infinite limit. Without lost of generality we can
consider the regulated risk process $X$ stopped at the ruin time.
If we assume then that there exists $\omega$ such that the limit $\tau^{c^*}$ of the $\tau^{c^*}_x$ is finite
then we get a contradiction though. Indeed, in this case we can always find $x$ large enough to get
\begin{equation*}
x+(\mu-c_0)\tau^{c^*}-\sum_{i=1}^{N^\omega_{\tau^{c^*}}}Y^\omega_i>0.
\end{equation*}
This contradiction completes the proof that the ruin time $\tau^{c^*}_x$ converges to infinity a.s. as $x\to\infty$.

By bounded convergence the quantity $\mathbb{E}_x \left[e^{-\beta\tau^{c^*}_x}\right]$ appearing in \eqref{ineq:NeededForLimitOfV} converges to zero as $x\to\infty$. Therefore $v(x)$ converges to $\frac{U(c_0)}{\beta}$ as $x\to\infty$.

Let us now prove that $v(x)$ is absolutely continuous.  Let $y>x\geq 0$. We denote by $c=(c_t)$ a strategy for the initial capital $y$. We take now the strategy $\overline{c}=(\overline{c}_t)$ that starts at initial reserve $x$, pays no dividends if $X_t^{\overline{c}}<y$ and follows strategy $c$ after reaching level $y$. This strategy $\overline{c}$ is of course admissible. In the event of no claims, the reserve process $X_t^{\overline{c}}$ reaches $y$ at time $t_0=\frac{y-x}{\mu}$. Since the probability of reaching $y$ before the first claim (i.e. $P(T_1>t_0)$) is $e^{-\lambda t_0}$ we get
\begin{equation*}
v(x)\geq v_{\overline{c}}(x)\geq P(T_1>t_0)e^{-\beta t_0}v(y)= e^{-(\lambda+\beta)\frac{y-x}{\mu}}v(y).
\end{equation*}
Since $v$ is increasing and bounded we get:
\begin{align}\label{eq:locLipineq}
0&\leq v(y)-v(x)\leq v(y)-e^{-(\lambda+\beta)\frac{y-x}{\mu}}v(y)\\&=v(y)\left(1-e^{-(\lambda+\beta)\frac{y-x}{\mu}}\right)\leq \frac{U(c_0)}{\beta}(\lambda+\beta)\frac{y-x}{\mu}.\notag
\end{align}
From that we obtain that $v$ is Lipschitz continuous on $[0,\infty)$, and consequently that $v$ is differentiable almost everywhere in $[0,\infty)$. Because of the Lipschitz continuity on compact sets, the above calculations give bounds for density $v'(x)$ which is therefore integrable on compact sets. This yields that $v$ is absolutely continuous, see for example Wheeden and Zygmund \cite{Wheeden}.
\end{proof}

\section{The Hamilton-Jacobi-Bellman equation}
In the proof of main theorem we will need lemma that constructs in a measurable way nearly optimal strategies for all 
initial states from a compact subset of $\mathbb{R}_+$. 
The proofs of this lemma and following theorem is based on the proofs of \cite[Theorem 2.32]{Schmidli} and \cite[Theorem 3.2]{EisenbergSchmidli}.
\begin{lem}\label{lem:NearlyOptimalStrategies}
For all $x\in [a,b]\subset [0,\infty)$ one can find in a measurable way  strategy $\hat{c}(x)$ such that $v_{\hat{c}(x)}(x)\geq v(x)-\varepsilon$.
\end{lem}
\begin{proof}
Choose $\varepsilon >0$ and  let $n\in\mathbb{N}$ be large enough such that:
\begin{equation}\label{eq:nLarge}
\frac{U(c_0)(\lambda+\beta)}{\beta\mu}\cdot\frac{b-a}{n}\leq\frac{\varepsilon}{2}.
\end{equation}
Let $x_k=\frac{k}{n}b+(1-\frac{k}{n})a$ for $0\leq k\leq n$. For each $x_k$ there is a strategy $\hat{c}(x_k)=\{\hat{c}(x_k)\}$ such that $v_{\hat{c}(x_k)}(x_k)>v(x_k)-\frac{\varepsilon}{2}$. Now we construct strategy for any $x\in [a,b]$. Take $x\in [a,b]$. Then there exist some $k_0$ such that $x_{k_0}\leq x<x_{k_0+1}$. For $x$ we choose the strategy $\hat{c}(x_{k_0})=\{\hat{c}(x_{k_0})\}$ i.e. the same strategy that for $x_{k_0}$. Then we have:
\begin{equation}
v_{\hat{c}(x)}(x)=v_{\hat{c}(x_{k_0})}(x)\geq v_{\hat{c}(x_{k_0})}(x_{k_0})\geq v(x_{k_0})-\frac{\varepsilon}{2}.
\end{equation}
From the inequalities \eqref{eq:locLipineq}, \eqref{eq:nLarge} and the definition of $x_{k_0}$ we get:
\begin{align*}
v(x_{k_0})&\geq  v(x)-\frac{U(c_0)(\lambda+\beta)}{\beta\mu}(x-x_{k_0})\nonumber\\
&\geq  v(x)-\frac{U(c_0)(\lambda+\beta)}{\beta\mu}\cdot\frac{b-a}{n}\nonumber\\
&\geq  v(x)-\frac{\varepsilon}{2}.
\end{align*}
This shows that for all $x\in [a,b]$ we can find in a measurable way a strategy $\hat{c}(x)$ such that $v_{\hat{c}(x)}(x)\geq v(x)-\varepsilon$.
\end{proof}
In the following theorem we derive Hamilton-Jacobi-Bellman equation and prove that value function $v(x)$ is smooth enough.
\begin{thm}\label{twr:DiffOfv}
If $c_{0}<\mu$, then the optimal value function $v(x)$ is continuously differentiable and fulfils Hamilton-Jacobi-Bellman equation:
\begin{equation}\label{eq:NoweHJBPelneInTheorem}
\sup_{0\leq c\leq c_0}\left\{(\mu-c)v'(x)-(\beta+\lambda)v(x)+U(c)+\lambda\int_{0}^{x}v(x-y)dF_Y(y)\right\}=0.
\end{equation}
\end{thm}
\begin{proof}

Since $v$ is Lipschitz continuous and increasing, hence $v$ is differentiable almost everywhere. Moreover, in the points where $v$ is not differentiable, $v$ is differentiable from the left and from the right.

Let $h>0$ and fix $c\in [0,c_0]$. Consider now the following strategy:

\begin{equation}\label{ConstantStrategyBeforeFirstClaim}
c_t=\begin{cases}
c, &\text{for $0\leq t\leq T_1\wedge h$,}\\
\hat{c}_{t-(T_1\wedge h)}(X_{T_1\wedge h}), &\text{for $t>T_1\wedge h$,}
\end{cases}
\end{equation}
where $\hat{c}(x)$ is a nearly optimal strategy defined in Lemma \ref{lem:NearlyOptimalStrategies}.
The first claim $T_1$ happens with density $\lambda e^{-\lambda h}$ and $T_1$ is larger than $h$ with probability $e^{-\lambda h}$. Thus:
\begin{align*}
v(x)&\geq e^{-\lambda h}\left[\int_0^h\hspace{-3pt} U(c)e^{-\beta t}dt+e^{-\beta h}v_{\hat{c}}(x+(\mu-c)h)\right]+\int_0^h\Big[\int_{0}^{t}\hspace{-3pt}U(c)e^{-\beta s}ds\\
&+e^{-\beta t}\int_0^{x+(\mu-c)t}v_{\hat{c}}(x+(\mu-c)t-y)dF_Y(y)\Big]\lambda e^{-\lambda t}dt\\
&\geq e^{-\lambda h}\left[\int_0^h U(c)e^{-\beta t}dt+e^{-\beta h}v(x+(\mu-c)h)\right]+\int_0^h\Big[\int_{0}^{t}U(c)e^{-\beta s}ds\\
&+e^{-\beta t}\int_0^{x+(\mu-c)t}v(x+(\mu-c)t-y)dF_Y(y)\Big]\lambda e^{-\lambda t}dt-\varepsilon
\end{align*}
The constant $\varepsilon$ is arbitrary. We therefore can let it tending to zero. Rearranging the terms and dividing by $h$ yields:
\begin{align}\label{eq:NierownoscZEpsilon}
0\geq & \frac{v(x+(\mu-c)h)-v(x)}{h}-\frac{1-e^{-(\beta+\lambda)h}}{h}v(x+(\mu-c)h)\\
&+e^{-\lambda h}\frac{1}{h}\int_0^h U(c)e^{-\beta t}dt+\frac{1}{h}\int_0^h\Big[\int_{0}^{t}U(c)e^{-\beta s}ds\notag\\
&+e^{-\beta t}\int_0^{x+(\mu-c)t}v(x+(\mu-c)t-y)dF_Y(y)\Big]\lambda e^{-\lambda t}dt.\notag
\end{align}
Now we choose a strategy $c(h)=(c_t(h))$ such that $v_c(x)\geq v(x)-h^2$. Note that $c_t(h)$ is just $c_t$ defined in \eqref{ConstantStrategyBeforeFirstClaim}. Let $a(t)=\mu t-\int_{0}^{t}c_s(h)ds$. Then:
\begin{align*}
&v(x)\leq h^2+v_c(x)=h^2+e^{-\lambda h}\left[\int_0^h U(c_t(h))e^{-\beta t}dt+e^{-\beta h}v_{c}\left(x+a(h)\right)\right]\\
&+\int_0^h\left[\int_{0}^{t}U(c_t(h))e^{-\beta s}ds+e^{-\beta t}\int_0^{x+a(t)}v_{c}\left(x+a(t)-y\right)dF_Y(y)\right]\lambda e^{-\lambda t}dt.
\end{align*}
Since $v_c(x)\leq v(x)$, it follows that:
\begin{align*}
&v(x)\leq h^2+e^{-\lambda h}\left[\int_0^h U(c_t(h))e^{-\beta t}dt+e^{-\beta h}v\left(x+a(h)\right)\right]\\
&+\int_0^h\left[\int_{0}^{t}U(c_t(h))e^{-\beta s}ds+e^{-\beta t}\int_0^{x+a(t)}\!\!v\left(x+a(t)-y\right)dF_Y(y)\right]\lambda e^{-\lambda t}dt.
\end{align*}
Rearranging the terms and dividing by $h$ yields:
\begin{align*}
0\leq &h+\frac{v\left(x+a(h)\right)-v(x)}{h}-\frac{1-e^{-(\beta+\lambda)h}}{h}v\left(x+a(h)\right)\\&+e^{-\lambda h}\frac{1}{h}\int_0^h U(c_t(h))e^{-\beta t}dt+\frac{1}{h}\int_0^h\Big[\int_{0}^{t}U(c_t(h))e^{-\beta s}ds\\&+e^{-\beta t}\int_0^{x+a(t)}v\left(x+a(t)-y\right)dF_Y(y)\Big]\lambda e^{-\lambda t}dt.
\end{align*}
Since $e^{-\beta t}\leq 1$ for $t\geq 0$, $U$ is nonnegative and strictly concave, from Jensen's inequality, it follows that:
\begin{equation*}
\frac{1}{h}\int_0^h U(c_t(h))e^{-\beta t}dt\leq \frac{1}{h}\int_0^h U(c_t(h))dt\leq U\left(\frac{1}{h}\int_0^h c_t(h)dt\right).
\end{equation*}
Therefore:
\begin{align}\label{eq:NierownoscZh}
0\leq & h+\frac{v\left(x+a(h)\right)-v(x)}{h}-\frac{1-e^{-(\beta+\lambda)h}}{h}v\left(x+a(h)\right)\\&+e^{-\lambda h}U\left(\frac{1}{h}\int_0^h c_t(h)dt\right)+\frac{1}{h}\int_0^h\Big[\int_{0}^{t}U(c_t(h))e^{-\beta s}ds\notag\\&+e^{-\beta t}\int_0^{x+a(t)}v\left(x+a(t)-y\right)dF_Y(y)\Big]\lambda e^{-\lambda t}dt.\notag
\end{align}
Note that:
\begin{equation*}
0\leq \int_{0}^{h}c_t(h)dt\leq \int_{0}^{h}c_0dt=c_0h.
\end{equation*}
Thus $a(h)\to 0$ as $h\to 0^+$. Since $v$ is continuous it follows that:
\begin{equation*}
\lim_{h\to 0^+}\frac{1-e^{-(\beta+\lambda)h}}{h}v\left(x+a(h)\right)=(\beta+\lambda)v(x)
\end{equation*}
and
\begin{equation*}
\lim_{h\to 0^+}\frac{1}{h}\int_0^h e^{-\beta t}\left(\int_0^{x+a(t)}\hspace{-25pt}v\left(x+a(t)-y\right)dF_Y(y)\right)\lambda e^{-\lambda t}dt=\lambda\int_{0}^{x}\hspace{-7pt}v(x-y)dF_Y(y).
\end{equation*}
Since $c_t$ are bounded, $U$ is nonnegative and continuous, and $0\leq e^{-\beta s}\leq 1$ for all $s\geq 0$, hence
\begin{equation*}
0\leq \frac{1}{h}\int_0^h\hspace{-6pt}\left(\int_{0}^{t}\hspace{-6pt}U(c_t(h))e^{-\beta s}ds\right)\lambda e^{-\lambda t}dt\leq \frac{1}{h}\int_{0}^{h}\hspace{-6pt}\left(\int_{0}^{t}\hspace{-6pt}U(c_0)ds\right)\lambda dt=\frac{\lambda U(c_0)}{2}h.
\end{equation*}
Therefore, we have:
\begin{equation*}
\lim_{h\to 0^+}\frac{1}{h}\int_0^h\left(\int_{0}^{t}U(c_t(h))e^{-\beta s}ds\right)\lambda e^{-\lambda t}dt=0.
\end{equation*}
Thus, with the exception of the second and fourth terms, the terms on the right-hand side of the inequality \eqref{eq:NierownoscZh} converge.

Consider now the second term. We have:
\begin{equation*}
\frac{v(x+a(h))-v(x)}{h}=\frac{v(x+a(h))-v(x)}{a(h)}\cdot \frac{a(h)}{h}.
\end{equation*}
Note that $\frac{1}{h}\int_{0}^{h}c_t(h)dt\in [0,c_0]$. Thus there exists a sequence $h_n\to 0^+$ such that:
\begin{equation*}
\lim_{h_n\to 0^+}\frac{1}{h_n}\int_{0}^{h_n}c_t(h_n)dt=\tilde{c}
\end{equation*}
for some $\tilde{c}\in [0,c_0]$. Note that $\tilde{c}$ might be random variable as for now. However, we will show later that there exists exactly one non-random value in $[0,c_0]$ which realizes the supremum in HJB equation \eqref{eq:NoweHJBzPrawej}.
Hence:
\begin{equation*}
\lim_{h_n\to 0^+}\frac{a(h_n)}{h_n}=\mu-\tilde{c}.
\end{equation*}
Since $U$ is continuous, it follows that
\begin{equation*}
\lim_{h_n\to 0^+}U\left(\frac{1}{h_n}\int_0^{h_n} c_t(h_n)dt\right)=U(\tilde{c}).
\end{equation*}
Recall that $c_0<\mu$, hence $a(h)>0$ and
\begin{equation*}
\lim _{h\to 0^+}\frac{v(x+a(h))-v(x)}{a(h)}=v'_+(x),
\end{equation*}
where $v'_+(x)$ means derivative from the right (which is finite as we mentioned above).
Thus, for the sequence $h_n\to 0^+$, chosen as above, we have from inequality \eqref{eq:NierownoscZh} that:
\begin{equation*}
0\leq (\mu-\tilde{c})v'_+(x)-(\beta+\lambda)v(x)+U(\tilde{c})+\lambda\int_{0}^{x}v(x-y)dF_Y(y).
\end{equation*}
Consider now inequality \eqref{eq:NierownoscZEpsilon}. Since $c\leq c_0<\mu$, it follows that $(\mu-c)h>0$ for all $c\in [0,c_0]$ and as a consequence:
\begin{equation*}
\lim_{h\to 0^+}\frac{v(x+(\mu-c)h)-v(x)}{(\mu-c)h}=v'_+(x).
\end{equation*}
Thus, for all $c\in [0,c_0]$, from inequality \eqref{eq:NierownoscZEpsilon} we have:
\begin{equation*}
0\geq (\mu-c)v'_+(x)-(\beta+\lambda)v(x)+U(c)+\lambda\int_{0}^{x}v(x-y)dF_Y(y).
\end{equation*}
It means that for $c=\tilde{c}$ we have:
\begin{equation*}
0=(\mu-\tilde{c})v'_+(x)-(\beta+\lambda)v(x)+U(\tilde{c})+\lambda\int_{0}^{x}v(x-y)dF_Y(y)
\end{equation*}
and $v'_+(x)$ fulfils Hamilton-Jacobi-Bellman equation:
\begin{equation}\label{eq:NoweHJBzPrawej}
\sup_{0\leq c\leq c_0}\left\{(\mu-c)v'_+(x)-(\beta+\lambda)v(x)+U(c)+\lambda\int_{0}^{x}v(x-y)dF_Y(y)\right\}=0
\end{equation}
in the case of $c_0<\mu$.
We can repeat the argument for any possible limit of
\begin{equation*}
\lim_{h_n\to 0^+}\frac{1}{h_n}\int_{0}^{h_n}c_t(h_n)dt,
\end{equation*}
leading to the same equation \eqref{eq:NoweHJBzPrawej}. This means that equation \eqref{eq:NoweHJBzPrawej} is satisfied for any $\omega\in \Omega$. 
Formally, the choice of $\tilde{c}$ dependents on chosen $\omega$ and the value of above limit may also depend on the sequence $h_n\to 0$.
However we will show below that there exists exactly one non-random value on $[0,c_0]$ which realizes the supremum in equation \eqref{eq:NoweHJBzPrawej}. 
In order to distinguish limits, we denote the corresponding limit by $\hat{c}$. Note that function $\zeta (c):=-cv'_+(x)+U(c)$ is strictly concave in $[0,c_0]$ because sum of strictly concave function and linear function is strictly concave. Since $\zeta$ is strictly concave, it follows that it has only one local maximum.
Therefore $\tilde{c}=\hat{c}$.
We showed then that there exists exactly one limit $\tilde{c}$ which realizes the supremum in equation \eqref{eq:NoweHJBzPrawej}.

Assume that $h$ is small enough such that $(x-(\mu-c)h)\wedge (x-a(h))>0$. Let us point out that $a(h)$ is dependent on initial state. Therefore the  $a(h)$ that is used from now on is different than the one defined in previous part of the proof. Moreover, since $c_t(h)$ depends on initial capital and we let $h\to 0$ in further part of the proof we do get use of measurable construction of nearly optimal strategies from Lemma \ref{lem:NearlyOptimalStrategies}. Starting with initial capital $x-(\mu-c)h$ inequality \eqref{eq:NierownoscZEpsilon} yields:
\begin{align}\label{eq:NierownoscZEpsilon2}
0\geq & \frac{v(x-(\mu-c)h)-v(x)}{-h}-\frac{1-e^{-(\beta+\lambda)h}}{h}v(x)\\
&+e^{-\lambda h}\frac{1}{h}\int_0^h U(c)e^{-\beta t}dt +\frac{1}{h}\int_0^h\Big[\int_{0}^{t}U(c)e^{-\beta s}ds\notag\\
&+e^{-\beta t}\int_0^{x-(\mu-c)(h-t)}v(x-(\mu-c)(h-t)-y)dF_Y(y)\Big]\lambda e^{-\lambda t}dt.\notag
\end{align}
Similarly as above, we conclude that:
\begin{equation}
0\geq (\mu-c)v'_{-}(x)-(\beta+\lambda)v(x)+U(c)+\lambda\int_0^x v(x-y)dF_Y(y)
\end{equation}
for all $c\in[0,c_0]$, and where $v'_{-}(x)$ means derivative from the left (which is finite as we mentioned above).
Starting with initial capital $x-a(h)$ inequality \eqref{eq:NierownoscZh} yields:
\begin{align}\label{eq:NierownoscZh2}
0\leq & h+\frac{v\left(x-a(h)\right)-v(x)}{-h}-\frac{1-e^{-(\beta+\lambda)h}}{h}v\left(x\right)\\&+e^{-\lambda h}U\left(\frac{1}{h}\int_0^h c_t(h)dt\right)+\frac{1}{h}\int_0^h\Big[\int_{0}^{t}U(c_t(h))e^{-\beta s}ds\notag\\&+e^{-\beta t}\int_0^{x-a(h)+a(t)}\!\!v\left(x-a(h)+a(t)-y\right)dF_Y(y)\Big]\lambda e^{-\lambda t}dt.\notag
\end{align}
Then there exists some $\hat{c}\in [0,c_0]$ such that:
\begin{equation}
0\leq (\mu-\hat{c})v'_{-}(x)-(\beta+\lambda)v(x)+U(\hat{c})+\lambda\int_0^x v(x-y)dF_Y(y).
\end{equation}
Hence $v'_{-}(x)$ satisfies Hamilton-Jacobi-Bellman equation:
\begin{equation}\label{eq:NoweHJBzLewej}
\sup_{0\leq c\leq c_0}\left\{(\mu-c)v'_{-}(x)-(\beta+\lambda)v(x)+U(c)+\lambda\int_{0}^{x}\hspace{-10pt}v(x-y)dF_Y(y)\right\}\hspace{-2pt}=0
\end{equation}
and similarly as above there exists exactly one limit $\hat{c}$ which realizes the supremum in equation \eqref{eq:NoweHJBzLewej}.

Now we prove that $v$ is differentiable (we show that $v'_+(x)=v'_{-}(x)$). We know that $v'_{+}$ and $v'_{-}$ satisfy equations \eqref{eq:NoweHJBzPrawej} and \eqref{eq:NoweHJBzLewej} respectively. We also know that there exist some values $\tilde{c}$ and $\hat{c}$ such that supremum in equations \eqref{eq:NoweHJBzPrawej} and \eqref{eq:NoweHJBzLewej} respectively are attained.
Thus:
\begin{equation}\label{eq:rownoscsupremow}
(\mu-\tilde{c})v'_+(x)+U(\tilde{c})=(\mu-\hat{c})v'_-(x)+U(\hat{c}).
\end{equation}
Since supremum in equation \eqref{eq:NoweHJBzLewej} is attained at $\hat{c}$, it follows that:
\begin{equation}\label{eq:nierownoscstrategi}
(\mu-\hat{c})v'_-(x)+U(\hat{c})\geq (\mu-\tilde{c})v'_-(x)+U(\tilde{c}).
\end{equation}
Then \eqref{eq:rownoscsupremow} combined with \eqref{eq:nierownoscstrategi} gives:
\begin{equation*}
(\mu-\tilde{c})v'_+(x)+U(\tilde{c})\geq (\mu-\tilde{c})v'_-(x)+U(\tilde{c}),
\end{equation*}
and therefore:
\begin{equation*}
(\mu-\tilde{c})(v'_+(x)-v'_-(x))\geq 0.
\end{equation*}
Since $\tilde{c}<\mu$, hence it immediately leads to
\begin{equation*}
v'_+(x)\geq v'_-(x).
\end{equation*}
On the other hand, since supremum in equation \eqref{eq:NoweHJBzPrawej} is attained at $\tilde{c}$, it follows that:
\begin{equation*}
(\mu-\tilde{c})v'_+(x)+U(\tilde{c})\geq (\mu-\hat{c})v'_+(x)+U(\hat{c}).
\end{equation*}
Then, similarly as above, we get that
\begin{equation*}
v'_-(x)\geq v'_+(x).
\end{equation*}
This means that $v$ is differentiable, and that $\tilde{c}=\hat{c}$.
Summarizing, we proved that there exist exactly one optimal value $c^*$ that simultaneously realizes supremum in equations \eqref{eq:NoweHJBzPrawej} and \eqref{eq:NoweHJBzLewej} and that the function $v$ is differentiable and satisfies Hamilton-Jacobi-Bellman equation:
\begin{equation}\label{eq:NoweHJBPelne}
\sup_{0\leq c\leq c_0}\left\{(\mu-c)v'(x)-(\beta+\lambda)v(x)+U(c)+\lambda\int_{0}^{x}v(x-y)dF_Y(y)\right\}=0
\end{equation}
in the case of $c_0<\mu$.
This completes the proof.
\end{proof}

\begin{thm}\label{twr:DiffOfv2}
If $c_{0}>\mu$ and $U$ is differentiable at $\mu$, then the optimal value function $v(x)$ is differentiable and fulfils the Hamilton-Jacobi-Bellman equation:
\begin{equation}\label{eq:NoweHJBPelne2InTheorem}
\sup_{0\leq c\leq c_0}\left\{(\mu-c)v'(x)-(\beta+\lambda)v(x)+U(c)+\lambda\int_{0}^{x}v(x-y)dF_Y(y)\right\}=0.
\end{equation}
If supremum in equation \eqref{eq:NoweHJBPelne2InTheorem} is not attained at $\mu$ then $v(x)$ is continuously differentiable.
\end{thm}
\begin{proof}
This proof is similar to the proof of Theorem \ref{twr:DiffOfv} with some significant modifications.

Since $v$ is Lipschitz continuous and increasing, hence $v$ is differentiable almost everywhere. Moreover, in the points where $v$ is not differentiable, $v$ is differentiable from the left and from the right.

Let $h>0$ and fix $c\in [0,c_0]$. If $x=0$, we suppose that $c\leq\mu$, and if $x>0$, we let $h$ be small enough such that $x+(\mu-c)h\geq 0$, i.e., ruin does not occur because of the dividend payments.  In the same way like in the mentioned proof we can derive inequality:
\begin{align}\label{eq:NierownoscZEpsilon3}
0\geq &\frac{v(x+(\mu-c)h)-v(x)}{h}-\frac{1-e^{-(\beta+\lambda)h}}{h}v(x+(\mu-c)h)\\
&+e^{-\lambda h}\frac{1}{h}\int_0^h U(c)e^{-\beta t}dt \notag+\frac{1}{h}\int_0^h\Big[\int_{0}^{t}U(c)e^{-\beta s}ds\\
&+e^{-\beta t}\int_0^{x+(\mu-c)t}v(x+(\mu-c)t-y)dF_Y(y)\Big]\lambda e^{-\lambda t}dt.\notag
\end{align}
From inequality \eqref{eq:NierownoscZEpsilon3} we have:
\begin{equation}\label{ineq:PlusCmMu}
0\geq (\mu-c)v'_+(x)-(\beta+\lambda)v(x)+U(c)+\lambda\int_{0}^{x}v(x-y)dF_Y(y)
\end{equation}
for $c<\mu$. Moreover,
\begin{equation}\label{ineq:MinusCwMu}
0\geq (\mu-c)v'_-(x)-(\beta+\lambda)v(x)+U(c)+\lambda\int_{0}^{x}v(x-y)dF_Y(y)
\end{equation}
for $c>\mu$ and
\begin{equation}\label{ineq:CrMu}
0\geq -(\beta+\lambda)v(x)+U(c)+\lambda\int_{0}^{x}v(x-y)dF_Y(y)
\end{equation}
for $c=\mu$.

Now we choose the strategy $c(h)=(c_t(h))$, such that $v_c(x)\geq v(x)-h^2$. Let $a(t)=\mu t-\int\limits_{0}^{t}c_s(h)ds$. Then in the same way like in the proof of Theorem \ref{twr:DiffOfv} we can derive inequality:
\begin{align}\label{eq:NierownoscZh3}
0\leq & h+\frac{v\left(x+a(h)\right)-v(x)}{h}-\frac{1-e^{-(\beta+\lambda)h}}{h}v\left(x+a(h)\right)\\&+e^{-\lambda h}U\left(\frac{1}{h}\int_0^h c_t(h)dt\right)+\frac{1}{h}\int_0^h\Big[\int_{0}^{t}U(c_t(h))e^{-\beta s}ds\notag\\&+e^{-\beta t}\int_0^{x+a(t)}\!\!v\left(x+a(t)-y\right)dF_Y(y)\Big]\lambda e^{-\lambda t}dt.\notag
\end{align}
Since $\frac{1}{h}\int_{0}^{h}c_t(h)dt\in [0,c_0]$, hence there exists at least one sequence $h_n\to 0^+$ such that:
\begin{equation*}
\lim_{h_n\to 0^+}\frac{1}{h_n}\int_{0}^{h_n}c_t(h_n)dt=\tilde{c}
\end{equation*}
and
\begin{equation*}
\lim_{h_n\to 0^+}\frac{a(h_n)}{h_n}=\mu-\tilde{c}.
\end{equation*}
From inequality \eqref{eq:NierownoscZh3} we have
\begin{equation}\label{ineq:PlusCTmMu}
0\leq (\mu-\tilde{c})v'_+(x)-(\beta+\lambda)v(x)+U(\tilde{c})+\lambda\int_{0}^{x}v(x-y)dF_Y(y)
\end{equation}
if $\tilde{c}<\mu$. Furthermore,
\begin{equation}\label{ineq:MinusCTwMu}
0\leq (\mu-\tilde{c})v'_-(x)-(\beta+\lambda)v(x)+U(\tilde{c})+\lambda\int_{0}^{x}v(x-y)dF_Y(y)
\end{equation}
if $\tilde{c}>\mu$ and
\begin{equation}\label{ineq:CTrMu}
0\leq -(\beta+\lambda)v(x)+U(\tilde{c})+\lambda\int_{0}^{x}v(x-y)dF_Y(y).
\end{equation}
if $\tilde{c}=\mu$.

Note that we do not know if $\tilde{c}$ is greater, less or maybe equal $\mu$. We do not know if there is just one $\tilde{c}$ either. It is possible that there is more than one $\tilde{c}$ which satisfy  inequality \eqref{ineq:PlusCTmMu} or \eqref{ineq:MinusCTwMu} or \eqref{ineq:CTrMu}. At the moment we know only that at least one inequality \eqref{ineq:PlusCTmMu} or \eqref{ineq:MinusCTwMu} or \eqref{ineq:CTrMu} is satisfied by at least one $\tilde{c}$.

Now we assume that $h$ is small enough such that $(x-(\mu-c)h)\wedge (x-a(h))>0$. Note that this $a(h)$ is different than $a(h)$ defined in previous part of the proof because we consider different initial capital below. Starting with initial capital $x-(\mu-c)h$, the inequality \eqref{eq:NierownoscZEpsilon3} gives:
\begin{align}\label{eq:NierownoscZEpsilon4}
0\geq & \frac{v(x-(\mu-c)h)-v(x)}{-h}-\frac{1-e^{-(\beta+\lambda)h}}{h}v(x)\\
&+e^{-\lambda h}\frac{1}{h}\int_0^h U(c)e^{-\beta t}dt +\frac{1}{h}\int_0^h\Big[\int_{0}^{t}U(c)e^{-\beta s}ds\notag\\
&+e^{-\beta t}\int_0^{x-(\mu-c)(h-t)}v(x-(\mu-c)(h-t)-y)dF_Y(y)\Big]\lambda e^{-\lambda t}dt.\notag
\end{align}
Similarly as above, we get \eqref{ineq:CrMu} for $c=\mu$ and the inequality
\begin{equation}\label{ineq:MinusCmMu}
0\geq (\mu-c)v'_{-}(x)-(\beta+\lambda)v(x)+U(c)+\lambda\int_0^x v(x-y)dF_Y(y)
\end{equation}
for $c<\mu$. Moreover,
\begin{equation}\label{ineq:PlusCwMu}
0\geq (\mu-c)v'_{+}(x)-(\beta+\lambda)v(x)+U(c)+\lambda\int_0^x v(x-y)dF_Y(y)
\end{equation}
for $c>\mu$.

Thus from \eqref{ineq:PlusCmMu}, \eqref{ineq:PlusCwMu} and \eqref{ineq:CrMu} we have that for all $c\in [0,c_0]$:
\begin{equation}\label{ineq:HJBPM}
0\geq (\mu-c)v'_{+}(x)-(\beta+\lambda)v(x)+U(c)+\lambda\int_0^x v(x-y)dF_Y(y)
\end{equation}
and from \eqref{ineq:MinusCwMu}, \eqref{ineq:MinusCmMu} and \eqref{ineq:CrMu} we have that for all $c\in [0,c_0]	$
\begin{equation}\label{ineq:HJBMM}
0\geq (\mu-c)v'_{-}(x)-(\beta+\lambda)v(x)+U(c)+\lambda\int_0^x v(x-y)dF_Y(y).
\end{equation}
This means that:
\begin{equation*}
\sup_{0\leq c\leq c_0}\left\{(\mu-c)v'_+(x)-(\beta+\lambda)v(x)+U(c)+\lambda\int_{0}^{x}v(x-y)dF_Y(y)\right\}\leq 0
\end{equation*}
and
\begin{equation*}
\sup_{0\leq c\leq c_0}\left\{(\mu-c)v'_-(x)-(\beta+\lambda)v(x)+U(c)+\lambda\int_{0}^{x}v(x-y)dF_Y(y)\right\}\leq 0.
\end{equation*}

Now, starting with initial capital $x-a(h)$ inequality \eqref{eq:NierownoscZh3} yields:
\begin{align}\label{eq:NierownoscZh4}
0\leq & h+\frac{v\left(x-a(h)\right)-v(x)}{-h}-\frac{1-e^{-(\beta+\lambda)h}}{h}v\left(x\right)\\&+e^{-\lambda h}U\left(\frac{1}{h}\int_0^h c_t(h)dt\right)+\frac{1}{h}\int_0^h\Big[\int_{0}^{t}U(c_t(h))e^{-\beta s}ds\notag\\&+e^{-\beta t}\int_0^{x-a(h)+a(t)}\!\!v\left(x-a(h)+a(t)-y\right)dF_Y(y)\Big]\lambda e^{-\lambda t}dt.\notag
\end{align}
Since $\frac{1}{h}\int_{0}^{h}c_t(h)dt\in [0,c_0]$, hence there exists at least one sequence $h_n\to 0^+$ such that:
\begin{equation*}
\lim_{h_n\to 0^+}\frac{1}{h_n}\int_{0}^{h_n}c_t(h_n)dt=\hat{c}
\end{equation*}
and
\begin{equation*}
\lim_{h_n\to 0^+}\frac{a(h_n)}{h_n}=\mu-\hat{c}.
\end{equation*}
Note here that $\hat{c}$ may be different than $\tilde{c}$.

From the inequality \eqref{eq:NierownoscZh4} we get \eqref{ineq:CTrMu} if $\hat{c}=\mu$ and
\begin{equation}\label{ineq:MinusCTmMu}
0\leq (\mu-\hat{c})v'_{-}(x)-(\beta+\lambda)v(x)+U(\hat{c})+\lambda\int_0^x v(x-y)dF_Y(y)
\end{equation}
if $\hat{c}<\mu$. Furthermore,
\begin{equation}\label{ineq:PlusCTwMu}
0\leq (\mu-\hat{c})v'_{+}(x)-(\beta+\lambda)v(x)+U(\hat{c})+\lambda\int_0^x v(x-y)dF_Y(y)
\end{equation}
if $\hat{c}>\mu$.

Thus from \eqref{ineq:PlusCTmMu}, \eqref{ineq:PlusCTwMu} and \eqref{ineq:CTrMu} there exists at least one value $c^*\in [0,c_0]$ such that
\begin{equation}\label{ineq:HJBPW}
0\leq (\mu-c^*)v'_{+}(x)-(\beta+\lambda)v(x)+U(c^*)+\lambda\int_0^x v(x-y)dF_Y(y)
\end{equation}
or from \eqref{ineq:MinusCTwMu}, \eqref{ineq:MinusCTmMu} and \eqref{ineq:CTrMu} there exists at least one value $c_*\in [0,c_0]$ such that
\begin{equation}\label{ineq:HJBMW}
0\leq (\mu-c_*)v'_{-}(x)-(\beta+\lambda)v(x)+U(c_*)+\lambda\int_0^x v(x-y)dF_Y(y).
\end{equation}
Note that above alternative is substantial.

Therefore, from inequalities \eqref{ineq:HJBPM}, \eqref{ineq:HJBPW} and \eqref{ineq:HJBMM} and \eqref{ineq:HJBMW}, we know that the following alternative:\\
either $v'_+(x)$ satisfies Hamilton-Jacobi-Bellman equation
\begin{equation}\label{eq:NoweHJBzPrawej2}
\sup_{0\leq c\leq c_0}\left\{(\mu-c)v'_+(x)-(\beta+\lambda)v(x)+U(c)+\lambda\int_{0}^{x}v(x-y)dF_Y(y)\right\}=0
\end{equation}
or
$v'_{-}(x)$ satisfies Hamilton-Jacobi-Bellman equation
\begin{equation}\label{eq:NoweHJBzLewej2}
\sup_{0\leq c\leq c_0}\left\{(\mu-c)v'_{-}(x)-(\beta+\lambda)v(x)+U(c)+\lambda\int_{0}^{x}v(x-y)dF_Y(y)\right\}=0.
\end{equation}
Note that we know only that alternative of this two sentences is true but at the moment we do not know if both sentences at the same time are true.

Note that function $\xi (c):=(\mu-c)v'_+(x)+U(c)$ is strictly concave in $[0,c_0]$, because sum of strictly concave function and linear function is strictly concave. Since $\zeta$ is strictly concave, it follows that it has only one local maximum. Therefore if the equation \eqref{eq:NoweHJBzPrawej2} is satisfied then there exists exactly one limit $c^*$ which realizes the supremum in this equation.
Similarly the function $\eta (c):=(\mu-c)v'_-(x)+U(c)$ is strictly concave in $[0,c_0]$. Therefore if the equation \eqref{eq:NoweHJBzLewej2} is satisfied then there exists exactly one limit $c_*$ which realizes the supremum in this equation.

We prove now that $v$ is differentiable. We show that $v'_+(x)=v'_-(x)$ in every possible case. Firstly we consider again inequalities \eqref{ineq:PlusCTmMu}, \eqref{ineq:MinusCTwMu} and \eqref{ineq:CTrMu}. As we mentioned above we don't know which of them are satisfied and how many values of $\tilde{c}$ there is. Note that all values $\tilde{c}$ (if there exists) realizes supremum in equations \eqref{eq:NoweHJBzPrawej2} or \eqref{eq:NoweHJBzLewej2}. We wrote above that there is exactly one value which realizes supremum in this equations (because of strict concavity of $\xi$ and $\eta$). Therefore there exists at most three values of $\tilde{c}$: $\tilde{c}_1<\mu$ or $\tilde{c}_2>\mu$ or $\tilde{c}_3=\mu$. We also know that at least one from this three values exists.

Similarly considering inequalities \eqref{ineq:MinusCTmMu}, \eqref{ineq:PlusCTwMu} and \eqref{ineq:CTrMu} we conclude that there exists at least one from at most three values of $\hat{c}$: $\hat{c}_1<\mu$ or $\hat{c}_2>\mu$ or $\hat{c}_3=\mu$, which realizes supremum in equations \eqref{eq:NoweHJBzPrawej2} or \eqref{eq:NoweHJBzLewej2}.

At the beginning we suppose that there exists exactly one value $\tilde{c}$ and exactly one value $\hat{c}$.
Now we consider all possible cases:\\\textbf{I.} $\tilde{c}_1<\mu \wedge\hat{c}_2>\mu$\\
In this case equations \eqref{ineq:PlusCTmMu} and \eqref{ineq:PlusCTwMu} are satisfied. This means that supremum in equation \eqref{eq:NoweHJBzPrawej2} is attained in $\tilde{c}_1$ and $\hat{c}_2$. Since supremum in equation  \eqref{eq:NoweHJBzPrawej2} may be attained in exactly one point thus we get $\tilde{c}_1=\hat{c}_2$ which leads to the contradiction. Thus this case is impossible to hold true.\\
\textbf{II.} $\tilde{c}_2>\mu \wedge\hat{c}_1<\mu$\\
In this case equations \eqref{ineq:MinusCTwMu} and \eqref{ineq:MinusCTmMu} are satisfied. This means that supremum in equation \eqref{eq:NoweHJBzLewej2} is attained in $\tilde{c}_2$ and $\hat{c}_1$. Since supremum in equation  \eqref{eq:NoweHJBzLewej2} may be attained in exactly one point thus we get $\tilde{c}_2=\hat{c}_1$ which leads to the contradiction. Thus this case is also impossible to hold true. Note that excluding case I. and II. leads to the corollary that $v'_+(x)$ fulfils equation \eqref{eq:NoweHJBzPrawej2} and $v'_-(x)$ fulfils equation \eqref{eq:NoweHJBzLewej2} (so now we know that both equations are satisfied).\\
\textbf{III.} $\tilde{c}_1<\mu \wedge\hat{c}_1<\mu$\\
This case was considered in Theorem \ref{twr:DiffOfv} (see \eqref{eq:rownoscsupremow}). We proved that $v'_+(x)=v'_-(x)$ in that case.\\
\textbf{IV.} $\tilde{c}_2>\mu \wedge\hat{c}_2>\mu$\\
Proof of this case is the same like in the case III. and leads to $v'_+(x)=v'_-(x)$.\\
\textbf{V.} $\tilde{c}_1<\mu \wedge\hat{c}_3=\mu$\\
In this case equations \eqref{ineq:PlusCTmMu} and \eqref{ineq:CTrMu} are satisfied. This means that supremum in equation \eqref{eq:NoweHJBzPrawej2} is attained in $\tilde{c}_1$ and $\mu$. Since supremum in equation  \eqref{eq:NoweHJBzPrawej2} may be attained in exactly one point thus we get $\tilde{c}_1=\mu$ which leads to the contradiction. Thus this case is impossible to hold true.\\
\textbf{VI.} $\tilde{c}_2>\mu \wedge\hat{c}_3=\mu$\\
\textbf{VII.} $\tilde{c}_3=\mu \wedge\hat{c}_1<\mu$\\
\textbf{VIII.} $\tilde{c}_3=\mu \wedge\hat{c}_2>\mu$\\
All these cases VI- VIII cannot hold for the same reasons as the ones given in the case V.\\
\textbf{IX.} $\tilde{c}_3=\mu \wedge\hat{c}_3=\mu$\\
In this case equation \eqref{ineq:CTrMu} is satisfied. But this means that supremum in equations \eqref{eq:NoweHJBzPrawej2} and \eqref{eq:NoweHJBzLewej2} is attained in $\mu$. From the definition of local maximum it follows that:
\begin{equation*}
U(\mu)\geq (\mu-c)v'_+(x)+U(c)
\end{equation*}
and
\begin{equation*}
U(\mu)\geq (\mu-c)v'_-(x)+U(c)
\end{equation*}
for all $c$ in some neighborhood of $\mu$. Since $U$ is concave, it follows that its derivatives from the left and right
exist. Consequently, above inequalities give:
\begin{equation*}
U'_+(\mu)\leq v'_+(x)\leq U'_-(\mu)
\end{equation*}
and
\begin{equation*}
U'_+(\mu)\leq v'_-(x)\leq U'_-(\mu).
\end{equation*}
Since $U$ is differentiable at $\mu$, hence $v'_+(x)=v'_-(x)$ in this case.

We considered all possible cases when there exists exactly one value $\tilde{c}$ and exactly one value $\hat{c}$ and obtained that $v'_+(x)=v'_-(x)$ in every possible case. Assume now that there exists more than one value of $\tilde{c}_1<\mu$ or $\tilde{c}_2>\mu$ or $\tilde{c}_3=\mu$ or there exists more than one value of $\hat{c}$: $\hat{c}_1<\mu$ or $\hat{c}_2>\mu$ or $\hat{c}_3=\mu$. This also leads to contradiction because there will be two or more different values which satisfy supremum in equation \eqref{eq:NoweHJBzPrawej2} or \eqref{eq:NoweHJBzLewej2}. We know that this is impossible since supremum in equations \eqref{eq:NoweHJBzPrawej2} and \eqref{eq:NoweHJBzLewej2} may be attained in exactly one point.

Therefore we proved that $v'_+(x)=v'_-(x)$ for each possible case.
This also means that the value $c^*$ (that realizes supremum in the equation \eqref{eq:NoweHJBzPrawej2}) and the value $c_*$ (that
realizes supremum in the equation \eqref{eq:NoweHJBzLewej2}) are equal to each other, i.e. $c^*=c_*$.
Summarizing, we proved that there exist exactly one optimal value $c^*$ that simultaneously realizes supremum in equations \eqref{eq:NoweHJBzPrawej2} and \eqref{eq:NoweHJBzLewej2} and that the function $v$ is differentiable and fulfils Hamilton-Jacobi-Bellman equation:
\begin{equation}\label{eq:NoweHJBPelne2}
\sup_{0\leq c\leq c_0}\left\{(\mu-c)v'(x)-(\beta+\lambda)v(x)+U(c)+\lambda\int_{0}^{x}v(x-y)dF_Y(y)\right\}=0
\end{equation}
in the case of $c_0>\mu$. From equation \eqref{eq:NoweHJBPelne2}, since $v$ and $U$ are continuous, it follows that $v$ is continuously differentiable if $c^{*}\neq \mu$. We proved that $v(x)$ is differentiable even for the points $x_0$ for which $c^{*}=\mu$. Note that derivative of a differentiable function never has a jump or removable discontinuity. Moreover, from the inequality \eqref{eq:locLipineq} it follows that $v'(x)$ is bounded and therefore cannot have infinite discontinuity. However we cannot exclude the situation that one or both limits $\lim_{x\to x_0-}v'(x),\lim_{x\to x_0+}v'(x)$ do not exist for the points $x_0$ for which $c^{*}=\mu$ and $v'(x)$ has essential discontinuity for the points $x_0$ for which $c^{*}=\mu$.
\end{proof}

From the fact that $v$ is differentiable and inequality \eqref{eq:locLipineq} we have following remark.

\begin{rem}
Function $v'$ is bounded and satisfies following inequality:
\begin{equation*}
0\leq v'(x)\leq\frac{U(c_0)}{\mu\beta}(\lambda+\beta).
\end{equation*}
\end{rem}

From now on we will assume that $U$ is differentiable and the Inada conditions are satisfied
i.e. $\lim\limits_{x\to 0}U'(x)=\infty$ and
$\lim\limits_{x\to\infty}U'(x)=0$.
Then we may calculate explicitly optimal strategy $c^*$.

\begin{rem}
If $U$ is differentiable and satisfy Inada conditions, then the optimal $c^*$ in equation \eqref{eq:NoweHJBPelne2InTheorem} is given by:
\begin{equation}
c^*(x)=
\begin{cases}\label{eq:OptimalStrategy}
(U')^{-1}(v'(x)), & \text{if } (U')^{-1}(v'(x))<c_0,\\
c_0, & \textrm{if } (U')^{-1}(v'(x))\geq c_0.
\end{cases}
\end{equation}
\end{rem}
Note that $c^*\geq 0$. Indeed, from Inada conditions we have that $(U')^{-1}(0)=\infty$ and $(U')^{-1}(\infty)=0$ and from the  strict concavity of $U$ we have that $U'$ and as a consequence $(U')^{-1}$ is strictly decreasing in $(0,\infty)$. Moreover $v$ is increasing. Hence $(U')^{-1}(v_x(x))\geq 0$ for all $x\geq 0$. Therefore $c^*\in [0,c_0]$.
This leads to the next remark.

\begin{rem}
If $U$ is differentiable and satisfy Inada conditions, then the optimal dividend strategy $(c^*_t)_{t\geq 0}$ is given by:
\begin{equation}\label{D:OptimalDividendStrategy}
c^*_t=
	\begin{cases}
	c^*(X_t), &\text{for $0\leq t<\tau$,}\\
	0, &\text{for $t\geq \tau$,}
	\end{cases}
\end{equation}
where $c^*$ is defined in equation \eqref{eq:OptimalStrategy}.
\end{rem}

\begin{thm}\label{T:VerificationTheorem}
Suppose that $\psi$ is a continuously differentiable, increasing, bounded, and nonnegative solution to \eqref{eq:NoweHJBPelneInTheorem} or \eqref{eq:NoweHJBPelne2InTheorem}. Then $\psi(x)=v(x)$ and an optimal dividend policy is given by Markovian strategy $c^*_t$ defined in \eqref{D:OptimalDividendStrategy}.
\end{thm}
\begin{proof}
This proof is based on the proof of \cite[Proposition 2.13]{AM}.
Function $\psi$ is nonnegative and continuously differentiable in $\mathbb{R}_{\geq 0}$ as a solution of equation \eqref{eq:NoweHJBPelneInTheorem} or \eqref{eq:NoweHJBPelne2InTheorem}. Since the function $e^{-\beta t}\psi(x)$ is continuously differentiable in $\mathbb{R}_{\geq 0}$, using the change of variables formula for finite variation process (see for instance \cite{Protter}), we can write
\begin{align}\label{eq:ItoFormula}
\psi(X_{t})e^{-\beta t}&-\psi(x)=\int_0^{t}\psi'(X_{s^-})e^{-\beta s}\mu ds-\beta\int_0^{t}\psi(X_{s^-})e^{-\beta s}ds\\
&-\int_0^{t}\psi'(X_{s^-})e^{-\beta s}dC_s+\sum_{\substack{X_{s^-}\neq X_{s}\\ s\leq t}}(\psi(X_{s})-\psi(X_{s^-}))e^{-\beta s}.\notag
\end{align}
Since $C_t=\int_0^tc_sds$ we have that
\begin{equation}\label{eq:FromAC}
\int_0^{t}\psi'(X_{s^-})e^{-\beta s}dC_s=\int_0^{t}\psi'(X_{s^-})e^{-\beta s}c_sds.
\end{equation}
On the other hand, $X_s\neq X_{s^-}$ only at the arrival of a claim, so by the compensation formula:
\begin{align}\label{eq:MtMartingale}
M_t =&\sum_{\substack{X_{s^-}\neq X_{s}\\ s\leq t}}(\psi(X_{s})-\psi(X_{s^-}))e^{-\beta s}\\
&-\lambda\int_0^te^{-\beta s}\int_0^{\infty}(\psi(X_{s^-}-y)-\psi(X_{s^-}))dF_Y(y)ds\notag
\end{align}
is a martingale with zero-expectation because:
\begin{equation*}
 0\leq \psi(X_s)\leq\max\limits_{y\in[0,x+\mu t]}\psi (y)<\infty,
\end{equation*}
for $s\leq t$. The upper bound inequality comes from the fact that $\psi$ is bounded and $X_t\leq x+\mu t$.
Therefore, we can combine \eqref{eq:ItoFormula}, \eqref{eq:FromAC} and \eqref{eq:MtMartingale} to obtain:
\begin{align*}
&\psi(X_{t})e^{-\beta t}-\psi(x)=\int_0^{t}\psi'(X_{s^-})e^{-\beta s}\mu ds-\beta\int_0^{t}\psi(X_{s^-})e^{-\beta s}ds+M_{t}\hspace{-1pt}\\
&-\int_0^{t}\psi'(X_{s^-})\hspace{-2pt}e^{-\beta s}c_sds+\lambda\int_0^{t}e^{-\beta s}\int_0^{\infty}(\psi(X_{s^-}-y)-\psi(X_{s^-}))dF_Y(y)ds\notag\\
=&\int_0^{t}\hspace{-6pt}\left(\psi'(X_{s^-})(\mu-c_s)-(\beta+\lambda)\psi(X_{s^-})+\lambda\int_0^{\infty}\hspace{-6pt}\psi(X_{s^-}-y)dF_Y(y)\right)e^{-\beta s}ds\\&+M_{t}.\notag
\end{align*}
Since $\psi$ fulfils equation \eqref{eq:NoweHJBPelneInTheorem} or \eqref{eq:NoweHJBPelne2InTheorem}, hence for an arbitrary strategy $(c_s)_{t\geq 0}$ we have:
\begin{equation}\label{I:WithMt}
\psi(X_{t})e^{-\beta t}-\psi(x)\leq -\int_0^t U(c_s)e^{-\beta s}ds+M_t.
\end{equation}
Taking expected values of both sides of above inequality gives:
\begin{equation}\label{I:BeforeLimit}
\psi (x) \geq \mathbb{E}_x \left(\psi(X_{t})e^{-\beta t}+\int_0^t U(c_s)e^{-\beta s}ds\right).
\end{equation}
Because $\psi (x)$ is bounded, $\lim_{t\to\infty}\mathbb{E}_x\left(\psi(X_{t})e^{-\beta t}\right)=0$. By the bounded convergence theorem,
\begin{equation}\label{I:AfterLimit}
\psi (x) \geq \mathbb{E}_x \left(\int_0^{\infty} U(c_s)e^{-\beta s}ds\right)=\mathbb{E}_x \left(\int_0^{\tau} U(c_s)e^{-\beta s}ds\right)=v_c(x),
\end{equation}
where in last equation we used the fact that $U(0)=0$ and $(c_s)_{t\geq 0}$ is admissible what means that there is no dividends after ruin occurs ($c_t=0$ for $t\geq \tau$). Since $\psi (x)\geq v_c(x)$ for an arbitrary strategy $(c_s)_{t\geq 0}$, it follows that $\psi (x)\geq v(x)$. If we take $(c^*_t)_{t\geq 0}$ defined in \eqref{D:OptimalDividendStrategy}, the the equality holds in \eqref{I:WithMt}, \eqref{I:BeforeLimit} and \eqref{I:AfterLimit}. Because $\psi (x)=v_{c^*}(x)$, it follows that $\psi (x)\leq v(x)$. This completes the proof.
\end{proof}

\section*{Acknowledgments}
This work is partially supported by the Ministry of Science and Higher Education of
Poland under the grant 2011/01/B/HS4/00982 (2012-2013).

The research of Sebastian Baran was supported by the Faculty of Finance, Cracow University of Economics, under grant \linebreak no. 158/WF-KM/02/2016/M/6158 for the research of young scientists.

\end{document}